\documentclass[lettersize,journal]{IEEEtran}
\usepackage{amsmath,amsfonts}
\usepackage{algorithmic}
\usepackage{algorithm}
\usepackage{array}
\usepackage[caption=false,font=normalsize,labelfont=sf,textfont=sf]{subfig}
\usepackage{cleveref}     
\usepackage{textcomp}
\usepackage{stfloats}
\usepackage{url}
\usepackage{verbatim}
\usepackage{graphicx}
\usepackage{subcaption}
\usepackage{cite}

\usepackage[table]{xcolor}
\usepackage{colortbl}
\usepackage{booktabs}
\usepackage{caption}

\begin{document}

\title{EEG-MSAF: An Interpretable Microstate Framework uncovers Default-Mode Decoherence in Early Neurodegeneration}

\author{Mohammad Mehedi Hasan$^{\star}$, Pedro G.~Lind$^{\star \dagger }$, Hernando Ombao$^{ \ddagger}$ , Anis Yazidi$^{\star \P}$ and Rabindra Khadka$^{\star \S }$. \thanks{ $^{*}$ Department of Computer Science, Oslo Metropolitan University, Oslo, Norway; $^{\ddagger}$ King Abdullah University of Science and Technology, Saudi Arabia; $^{\P}$ Department of Informatics, University of Oslo, Norway;  $^{\dagger}$ Kristiania University of Applied Sciences, Norway; {$^{\S}$Corresponding author. Email: \texttt{rabindra@oslomet.no}} }}
\date{}



\maketitle

\begin{abstract}
 Dementia (DEM) is a growing global health challenge, emphasizing the need for early and accurate diagnostic methods. Electroencephalography (EEG) offers a promising non-invasive approach to detecting subtle neurological changes, yet conventional methods often fail to capture the brain activity's transient and complex nature. To this end, we introduce a EEG Microstate Analysis Framework (EEG-MSAF), an end-to-end framework that leverages EEG microstates-discrete, quasi-stable topographical patterns to identify DEM-related biomarkers, and feature ranking to identify key neural biomarkers distinguishing DEM, mild cognitive impairment (MCI), and healthy controls, i.e., normal cognition (NC). Our approach encompasses three key stages: (1) automated extraction of microstates' features, (2) classification using machine learning (ML) algorithms to distinguish between DEM, MCI, and NC, and (3) feature ranking via Shapley Additive Explanations (SHAP) to identify the most relevant microstates' features contributing to disease differentiation. Experiments on two independent EEG datasets are presented in detail. One is the publicly available Chung-Ang University EEG (CAUEEG) dataset, and the other is a clinical cohort from Thessaloniki Hospital.
 These two datasets showcase robust performance and generalizability of the EEG-MSAF. On the CAUEEG dataset, our EEG-MSAF-SVM model achieved the state-of-the-art accuracy of $89\%\pm0.01$, outperforming the deep learning (DL) baseline CEEDNET by over 19.3\%. Likewise, on the Thessaloniki dataset, our model achieved $95\%\pm0.01$ accuracy, matching the performance of EEGConvNeXt. Moreover, our SHAP analysis highlights mean correlation and occurrence as the most informative microstate metrics: disruption of microstate C (salience/attention network) emerges as the dominant marker of DEM, while microstate F, a newly described default-mode pattern, ranks among the top predictors for both MCI and DEM. These findings position microstate F as a practical, early EEG biomarker of the anterior default mode network (DMN). By combining performance, generalizability, and interpretability, our framework not only advances EEG-based DEM diagnosis but also offers insight into the reorganization of brain dynamics across the cognitive spectrum. 
\end{abstract}


\section{Introduction}

Dementia (DEM) is a growing global health crisis with an increasing number of cases worldwide~\cite{livingston2020dementia}. This has largely impacted individuals, families, healthcare systems, and the economy~\cite{wimo2013worldwide}.  Alzheimer’s disease (AD) is the most common form of DEM, and early detection of mild cognitive impairment (MCI), which often precedes AD, is crucial as timely interventions targeting modifiable risk factors can potentially delay or even prevent the progression to DEM~\cite{van2023towards,rafii2023detection}. 

Electroencephalography (EEG) has emerged as a practical and non-invasive neuroimaging modality for detecting early neurophysiological changes in DEM~\cite{adamis2005utility}. Due to its excellent temporal resolution and low cost, EEG is particularly suited for longitudinal cognitive monitoring and scalable clinical deployment. Among EEG-based approaches, microstate analysis has gained traction for characterizing large-scale brain dynamics~\cite{zanesco2021meditation,metzger2024functional,kim2024unveiling}. EEG microstates are short-lived (80–120 ms), quasi-stable topographical patterns that are believed to reflect coordinated activity of resting-state neural networks~\cite{lehmann1987eeg,michel2018eeg,britz2010bold}. 

Each canonical microstate (A, B, C, F) has been functionally linked to distinct neural systems: Microstate A is associated with the auditory network and phonological processing; Microstate B with the visual network and visual attention; Microstate C with the salience network, supporting cognitive control and decision-making, 
and Microstate F, associated with the anterior default mode network (DMN), play a role in personally significant information processing, mental simulations, and theory of mind~\cite{brechet2019capturing, tarailis2024functional}. 

Several studies have suggested that microstates A, B, C, D, and F correspond to temporal, occipital, medial temporal, frontal lobe networks, and bilateral activity in medial prefrontal cortex, respectively~\cite{britz2010bold,tarailis2024functional,brechet2019capturing}. Alterations in the temporal parameters of these microstates—such as mean duration, occurrence rate per second, and time coverage—have been linked to various neurological disorders, including AD and MCI~\cite{musaeus2019microstates, musaeus2020changes, strik1997decreased}. These functional associations make microstates a powerful lens for interpreting disrupted brain network dynamics in cognitive decline.

 Braak et al. ~\cite{braak1991neuropathological} reported that early amyloid deposition begins in the isocortex, particularly in the basal portions of the temporal, occipital, and frontal lobes. These spatial patterns map closely onto the cortical origins of Microstates A (temporal), B (occipital), and D (frontoparietal), supporting the hypothesis that abnormal increases in the activity or duration of these microstates may reflect early, region-specific neuropathological changes. There is also evidence that specific microstate classes (e.g., microstate C or D) are affected in patients exhibiting cognitive decline~\cite{yan2024abnormal, lassi2023degradation}. The study by Musaeus et al.~\cite{musaeus2019microstates} reported significantly reduced time coverage and occurrence in AD patients. Another study by Lassi et al.~\cite{lassi2023degradation} found that microstate topographies in AD patients displayed higher discriminatory power than traditional spectral or network-based features. These findings highlight the utility of microstate-based descriptors as biologically interpretable and disease-relevant EEG biomarkers.

Recent work has explored integrating microstate features with ML algorithms to transform these insights into actionable diagnostic tools. Traditional ML models 
have proven effective for classifying cognitive states when trained on carefully engineered features~\cite{wu2025unveiling, song2022eeg, yang2024resting}. Compared to deep learning (DL) models, which typically require large datasets and offer limited interpretability, these traditional algorithms are more transparent, lightweight, and better suited for offline analysis and clinical applications. Recent studies have also shown that tree-based models often outperform DL on structured, low-dimensional datasets where domain knowledge can be effectively encoded through feature extraction~\cite{grinsztajn2022tree,shwartz2022tabular}. In this work, we adopt three traditional ML models, namely support vector machines (SVM), Random Forests (RFs), and extreme gradient boosting (XGB), specifically due to the tabular nature of the engineered microstate features used as input.

However, despite the practical advantages of traditional models, a critical limitation remains; their predictions often lack interpretability. Clinical deployment requires not only accurate predictions but also clear, explainable insights into the model's decision-making process. In this context, explainable AI (XAI) techniques such as SHapley Additive exPlanations (SHAP)~\cite{lundberg2017shap, winter2002shapley} have been adopted to provide feature-level attributions and enhance trustworthiness. SHAP has been successfully applied in recent studies to rank and quantify the importance of EEG-derived features in AD classification tasks~\cite{vimbi2024interpreting}, thereby supporting both model validation and clinical reasoning.

\begin{figure*}[ht]
    \centering
    \includegraphics[width=0.98\textwidth]{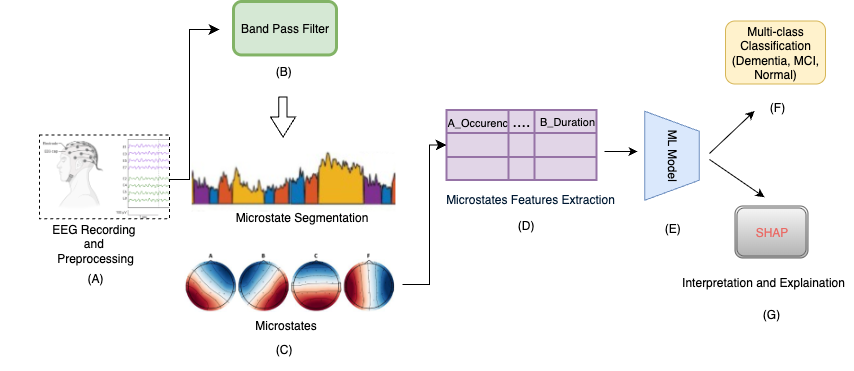}
    \caption{\textbf{Schematic of EEG-MSAF for DEM classification.} (A) EEG signals are recorded and preprocessed, followed by (B) band-pass filtering, allowing selection of specific EEG frequency bands of interest. (C) Microstate segmentation is performed to identify canonical microstates. (D) Features such as occurrence, duration, and coverage are extracted for each microstate and stored in tabular format. (E) The features are input into three traditional ML models (SVM, RFs, XGB). (F) The trained model performs multi-class classification of DEM, MCI, and NC. (G) SHAP is used for post hoc interpretation, providing feature-level explanations to support model transparency and clinical insight. }
    \label{fig:schema}
     \vspace{2em}
\end{figure*}

Despite the progress in microstate analysis, ML, and explainability, the literature still lacks an integrated framework that combines these components in a coherent and scalable pipeline. Most studies focus on one aspect, microstate computation without classification~\cite{grieder2016discovering}, classification without interpretability~\cite{li2024data}, or explainability applied to heterogeneous EEG features~\cite{morabito2023explainable}. Moreover, the majority of existing work deals with binary classification (e.g., AD vs. healthy), leaving multi-class classification (NC vs. MCI vs. DEM) underexplored. This is particularly important given the clinical need to distinguish MCI as an intermediate and potentially reversible stage.

To address these gaps, we present the EEG Microstate Analysis Framework (EEG-MSAF), a novel and interpretable ML framework for EEG-based classification of cognitive impairment, specifically targeting the early detection of DEM (see Figure~\ref{fig:schema}). Our approach is designed to be modular, clinically meaningful, and scalable across datasets. We make the following key contributions:

\begin{itemize}
    \item \textbf{End-to-end pipeline for interpretable DEM classification:} We propose the EEG Microstate Analysis Framework (EEG-MSAF), a unified framework that integrates EEG microstate feature extraction, multi-class classification (NC, MCI, and DEM), and post hoc explainability using SHAP values. To our knowledge, this is the first study to bring together these components in a coherent, end-to-end system evaluated on a clinical EEG dataset.
    
    \item \textbf{ EEG microstates feature extraction:} We extract meaningful microstate features, namely duration, occurrence, and coverage-from resting-state EEG, enabling the model to leverage interpretable neural dynamics associated with cognitive decline. Unlike prior work focused on raw signal learning, our feature-based approach provides transparency and relevance for clinical application.
    
    \item \textbf{State-of-the-art results on the CAUEEG dataset:} We evaluate our framework on the publicly available CAUEEG dataset, which includes recordings from individuals with NC, MCI, and DEM. Our model, EEG-MSAF-SVM achieves state-of-the-art performance in multi-class classification. We further validate the approach on a smaller DEM dataset from the General Hospital of Thessaloniki, demonstrating its robustness and generalizability.

    \item \textbf{Insightful explainability through SHAP analysis:} We apply SHapley Additive exPlanations (SHAP) to quantify the contribution of each microstate feature to model predictions. This allows us to surface neurophysiological patterns most indicative of cognitive impairment, addressing a critical gap in model interpretability.

    \item \textbf{Identification of microstate F as an early biomarker:} SHAP consistently ranks \texttt{F\_mean\_corr}, \texttt{F\_occurrences}, and \texttt{F\_mean\_dur} among the most influential features for both MCI and DEM, pointing to early anterior DMN disruption and establishing microstate F as a practical EEG marker.

\end{itemize}

Section~\ref{method} introduces the proposed framework in detail, describing the methodology for EEG microstate segmentation, feature extraction, and the architecture of the classification and explainability pipeline. Section~\ref{exp} details the experimental setup. Section~\ref{results} presents an extensive empirical evaluation of the framework. 
In Section~\ref{discuss}, we discuss the results, limitations, and future work. Finally, Section~\ref{conclusion} concludes the paper with a summary of findings.


\section{Methodology}\label{method}


\subsection{EEG Data from Chung-Ang University  (CAUEEG)}

The CAUEEG dataset~\cite{kim2023deep} is a publicly available resting-state EEG dataset for DEM research. It includes 21-channel recordings, recorded at a sampling frequency of 200 Hz using the international 10–20 system, along with ECG and photic stimulation channels. The dataset was collected at Chung-Ang University Hospital, South Korea. It comprises EEG data from a total of 1,155 participants, categorized into three clinically diagnosed groups: NC, MCI, and DEM. Diagnostic labels in the CAUEEG dataset were assigned based on established clinical criteria. DEM diagnoses followed NINCDS-ADRDA and DSM-IV guidelines~\cite{dubois2007research, first2002dsm}. MCI subjects met criteria that included memory complaints, intact daily functioning, objective cognitive deficits across multiple domains, and a Clinical Dementia Rating (CDR) of 0.5~\cite{ahn2010seoul, jahng2015constructing}. Normal controls (NC) had no cognitive impairments (within 1.0 SD of normative scores) and intact daily functioning. The dataset comprises 459 recordings labeled as NC, 416 as MCI, and 311 as DEM. The mean age of the participants is 70.77 years with a standard deviation of 9.90 years. There is a moderate imbalance of gender distribution, with approximately 60 males per 100 females (see Table~\ref{tab:patient_statistics}). To assess whether age distributions differed significantly across diagnostic groups (NC, MCI, and DEM), we also performed a non-parametric Kruskal–Wallis H-test~\cite{macfarland2016kruskal} given the non-normality of age distributions in each group, as indicated by the Shapiro–Wilk test~\cite{hanusz2016shapiro}. We found a statistically significant difference in age distributions across the diagnostic groups ($H = 295.49, p < 0.05$), suggesting that age varies meaningfully between healthy controls, individuals with MCI, and those with DEM. 

The CAUEEG dataset offers several advantages for developing ML models for DEM classification: it is balanced across cognitive stages, includes sufficient temporal resolution for microstate analysis, and reflects real-world clinical heterogeneity. 

\begin{table}[t]
\definecolor{headercolor}{RGB}{224, 217, 202} 
\definecolor{rowgray}{gray}{0.95}             
\renewcommand{\arraystretch}{1.3} 
\centering
\caption{Summary statistics for NC, MCI, DEM group. The total number of males and females is derived from the given ratio in the CAUEEG dataset~\cite{kim2023deep}.}
\label{tab:patient_statistics}
\resizebox{\columnwidth}{!}{%
\begin{tabular}{lccccc}
\rowcolor{headercolor}
\textbf{Group} & \textbf{Mean Age} & \textbf{Age Std.} & \textbf{ Female} & \textbf{Male} & \textbf{ Total} \\
\rowcolor{rowgray}
NC   & 65.10  & 9.48 & 172  & 287 & 459   \\
MCI      & 73.70 & 7.89 & 157 &  260 & 417  \\
\rowcolor{rowgray}
DEM &76.63  & 8.07 & 117 & 194 & 311   \\
\end{tabular}
}
\end{table}

\subsection{EEG data from the General Hospital of Thessaloniki}

To evaluate the generalizability of our proposed framework, we utilized a secondary dataset comprising resting-state EEG recordings from patients at the General Hospital of Thessaloniki~\cite{ntetska2025complementary}. It provides recordings from individuals diagnosed with AD, frontotemporal dementia (FTD), and healthy controls (CN). EEG recordings were acquired using a 19-channel cap configured according to the international 10–20 electrode placement system. Data were sampled at 500 Hz and collected under resting-state, eyes-open conditions using a referential montage (Cz reference). The dataset includes 36 participants with AD, 23 with FTD, and 29 healthy controls. 
The average participant age ranged from 63.6 to 67.9 years across groups.



\subsection{Preprocessing}
\label{preproc}

For the CAUEEG dataset, we selected 19 EEG channels, and each EEG recording was band-pass filtered between 0.5 Hz and 40 Hz using a finite impulse response (FIR)~\cite{widmann2015digital} filter to eliminate slow signal drifts and high-frequency noise from the data. We applied channel-wise z-score standardization (mean = 0, standard deviation = 1) followed by average referencing by projection to reduce channel-wise variability and improve signal consistency across electrodes. To reduce edge-related artifacts, we cropped the first and last minute of each recording before analysis.

For the General Hospital of Thessaloniki dataset, we utilized the preprocessed EEG signals provided, which were downsampled from 500 Hz to 100 Hz. The original preprocessing pipeline included band-pass filtering (0.5–40 Hz), artifact correction via Artifact Subspace Reconstruction (ASR), and Independent Component Analysis (ICA) using the RunICA algorithm. Components classified as eye or muscle artifacts were automatically rejected using the ICLabel plugin in EEGLAB. Further, we applied notch filtering and Laplacian spatial filtering~\cite{perrin2007scalp}, often referred to as Current Source Density (CSD) or Surface Laplacian (SL), to improve the topography of the microstates and reduce volume conduction.   

\subsection{Microstate Segmentation}

Microstates are brief, quasi-stable EEG topographies that typically persist for 60–120 ms before rapidly transitioning to another configuration~\cite{michel2018eeg}. To identify the most recurrent spatial patterns, we applied a data-driven microstate segmentation procedure based on global field power (GFP) and modified $k$-means clustering, following established protocols~\cite{pascual1995segmentation, koenig2002millisecond}.

For each subject, we first computed the Global Field Power (GFP), defined as the standard deviation of all electrode potentials at each time point (see Equation~\ref{gfp}), and identified its peaks to capture moments of maximal topographic stability. GFP quantifies the overall strength of the electrical field across the scalp, so a high GFP value suggests a strong well-defined electrical field and a low GFP indicates a weak or flat field. We applied a modified $k$-means clustering algorithm at the GFP topographies to extract subject-level microstate maps. We determined the optimal number of clusters based on the global explained variance (GEV) criterion. GEV quantifies how much a given microstate map explains the varinace of the original EEG data. 

Consistent with prior~\cite{michel2018eeg,tarailis2024functional}, we found that four microstate classes (labeled A, B, C, F) provided an interpretable model of the data. Then the individual topographies from each subject in the group are collected, and the second $k$-means clustering algorithm is applied for the group-level analysis. The resulting fitted clustering algorithm is used to predict the segmentation on each subject´s EEG recording. This process ensures the backfitting of group-level maps to each recording. All microstate segmentation and back-fitting procedures were implemented using the pycrostate library~\cite{ferat2022pycrostates}.
After this, we proceed to extract microstate features: 
\begin{equation}
GFP(t) = \sqrt{\frac{1}{K} \sum_{i=1}^{K} \left(V_i(t) - V_{\text{mean}}(t)\right)^2}
\label{gfp}
\end{equation}
where $V_i(t)$ is the potential at time $t$ for electrode $i$, $V_{\text{mean}}(t)$ is the average potential across all electrodes at time $t$, and $K$ is the total number of electrodes.

\subsection{Feature Extraction}

Following microstate segmentation and backfitting, we extracted a comprehensive set of features characterizing each microstate's temporal and spatial dynamics. Each microstate class (A, B, C, F) was described using five standard metrics widely adopted in the microstate literature~\cite{michel2018eeg,koenig2002millisecond}. These features were computed from the backfitted microstate sequence of each subject, resulting in individualized microstate profiles suitable for downstream ML analysis.

Specifically, the following features were computed for each microstate:
\begin{itemize}
    \item \textbf{Global Explained Variance (GEV)}: The proportion of total variance in the EEG signal explained by a given microstate. GEV quantifies the correlation between the chosen microstate topographic map and the topographies at each time point~\cite{pascual1995segmentation}. As shown in Equation~\ref{eq:GEV}, the GEV is expressed as the sum of squared spatial correlations between the instantaneous EEG topography at each time point and its corresponding microstate map, weighted by the Global Field Power (GFP) at that time point, normalized by the total GFP of the data.

    \begin{equation}
    \text{GEV}_k = \frac{\sum_{t=1}^{T} \left( \text{GFP}(t)^2 \cdot r_k(t)^2 \right)}{\sum_{t=1}^{T} \text{GFP}(t)^2}
    \label{eq:GEV}
    \end{equation}
    where  \( \text{GEV}_k \) is the Global Explained Variance of microstate class \( k \), \( \text{GFP}(t) \) is the Global Field Power at time point \( t \), \( r_k(t) \) is the spatial correlation between the EEG map at time \( t \) and the topographic map of microstate class \( k \), \( T \) is the total number of time points.
    
    \item \textbf{Mean Correlation (mean\_corr)}: The average spatial correlation between the microstate template and time point assigned to that microstate~\cite{koenig2002millisecond}, as shown in Equation~\ref{eq:mean_corr}.

    \begin{equation}
    \text{mean\_corr}_k = \frac{1}{N_k} \sum_{t \in T_k} \text{corr}(\mathbf{x}_t, \mathbf{ms}_k)
    \label{eq:mean_corr}
    \end{equation}
    
    \noindent where \( T_k = \{ t \, | \, s(t) = k \} \) is the set of time points assigned to microstate \( k \), \( N_k \) is the number of such time points (i.e., \( N_k = |T_k| \)), \( \text{corr}(\mathbf{x}_t, \mathbf{ms}_k) \) is the Pearson correlation between the EEG topography at time \( t \) and the microstate map of class $k$ (\( \mathbf{ms}_k \)).
    
    \item \textbf{Time Coverage (time\_cov)}: The fraction of the total EEG recording time during which a microstate was active~\cite{koenig2002millisecond} (see Equation~\ref{eq:time_cov}), indicating its temporal dominance.
    
    \begin{equation}
    \text{Time Coverage}_k = \frac{T_k}{T_{\text{total}}}
    \label{eq:time_cov}
    \end{equation}
    
    \noindent where \( T_k \) is the total duration (in samples or seconds) that microstate \( k \) is active, \( T_{\text{total}} \) is the total duration of the EEG recording.

    \item \textbf{Mean Duration (mean\_dur)}: The average duration (in milliseconds) of continuous segments assigned to the microstate.
    \begin{equation}
     \overline{D} = \frac{1}{N} \sum_{i=1}^{N} d_i
    \label{eq:mean_duration}
    \end{equation}

    \noindent where \( \overline{D} \) is the \textbf{mean duration} of a given microstate class, \( d_i \) denotes the duration (in milliseconds or time points) of the \( i^{\text{th}} \) microstate segment,  \( N \) is the total number of microstate segments observed for that class.

    \item \textbf{Occurrence Rate (occurrence)}: The number of times a microstate appeared per second, expressed as segments per second, providing a measure of frequency.
    
    \begin{equation}
    R = \frac{N}{T}
    \label{eq:occ}
    \end{equation}

    \noindent where \( R \) is the \textbf{occurrence rate}, defined as the number of times a given microstate appears per second, \( N \) is the total number of microstate segments identified for that class, \( T \) is the total duration of the EEG recording (in seconds).

\end{itemize}

For each subject, we extracted these five features for all four microstate classes (A, B, C, F), resulting in a total of 20 microstate-derived features. Additionally, we computed the subject-level Global Field Power (GFP) as an aggregate measure of synchrony across the entire brain network, bringing the total number of features to 21. The features were stored in a structured tabular format, with each row representing a subject and each column representing a specific feature. An overview of the extracted features is presented in Table~\ref{tab:microstate_features}.

\begin{table}[ht]
\centering
\caption{List of Extracted Microstate Features}
\label{tab:microstate_features}

\definecolor{headercolor}{RGB}{224, 217, 202} 
\definecolor{rowgray}{gray}{0.95}             

\begin{tabular}{|c|l|l|}
\rowcolor{headercolor}
\hline
\textbf{No} & \textbf{Feature} & \textbf{Comments} \\
\hline
\rowcolor{white}
1  & A\_gev         & Global explained variance of microstate A \\
\rowcolor{rowgray}
2  & A\_meancorr    & Mean correlation of microstate A \\
\rowcolor{white}
3  & A\_occurrence  & Occurrence of microstate A \\
\rowcolor{rowgray}
4  & A\_timecov     & Time coverage of microstate A \\
\rowcolor{white}
5  & A\_meandur     & Mean duration of microstate A \\
\rowcolor{rowgray}
6  & B\_gev         & Global explained variance of microstate B \\
\rowcolor{white}
7  & B\_meancorr    & Mean correlation of microstate B \\
\rowcolor{rowgray}
8  & B\_occurrence  & Occurrence of microstate B \\
\rowcolor{white}
9  & B\_timecov     & Time coverage of microstate B \\
\rowcolor{rowgray}
10 & B\_meandur     & Mean duration of microstate B \\
\rowcolor{white}
11 & C\_gev         & Global explained variance of microstate C \\
\rowcolor{rowgray}
12 & C\_meancorr    & Mean correlation of microstate C \\
\rowcolor{white}
13 & C\_occurrence  & Occurrence of microstate C \\
\rowcolor{rowgray}
14 & C\_timecov     & Time coverage of microstate C \\
\rowcolor{white}
15 & C\_meandur     & Mean duration of microstate C \\
\rowcolor{rowgray}
16 & F\_gev         & Global explained variance of microstate F \\
\rowcolor{white}
17 & F\_meancorr    & Mean correlation of microstate F \\
\rowcolor{rowgray}
18 & F\_occurrence  & Occurrence of microstate F \\
\rowcolor{white}
19 & F\_timecov     & Time coverage of microstate F \\
\rowcolor{rowgray}
20 & F\_meandur     & Mean duration of microstate F \\
\rowcolor{white}
21 & gfp            & Global field power  \\
\hline
\end{tabular}
\end{table}

\subsection{Classification Models}

Given the tabular structure and moderate dimensionality of the extracted microstate features, we adopt traditional ML models that are well-suited for structured data and offer robust performance with relatively limited sample sizes. Specifically, we employ separately SVM, RFs, and XGB to perform multi-class classification of subjects into NC, MCI, or DEM. These models are widely used in biomedical data analysis and provide competitive accuracy along with varying degrees of interpretability.

\subsubsection{Support Vector Machine (SVM)}

SVMs are a class of supervised learning algorithms that separate data points belonging to different classes by maximizing the margin between them~\cite{cortes1995support}. In their original formulation, SVMs are designed for binary classification. However, they can be effectively extended to handle multi-class problems using strategies such as \textit{one-vs-rest} (OvR) and \textit{one-vs-one} (OvO), both of which are supported by standard libraries like \texttt{scikit-learn}.

In this work, we adopt the one-vs-rest (OvR) strategy for multi-class classification, where $K$ separate binary classifiers are trained, one for each class against all others. During inference, each classifier outputs a decision function, and the class with the highest score is selected:
\[
\hat{y} = \arg\max_{k \in \{1, \ldots, K\}} f_k(\mathbf{x}),
\]
where $f_k(\mathbf{x})$ denotes the decision function of the $k$-th binary SVM.

Each binary SVM solves the following convex optimization problem:
\[
\min_{\mathbf{w}, b, \boldsymbol{\xi}} \frac{1}{2} \|\mathbf{w}\|^2 + C \sum_{i=1}^{N} \xi_i
\]
\[
\text{subject to:} \quad y_i(\mathbf{w} \cdot \phi(\mathbf{x}_i) + b) \geq 1 - \xi_i, \quad \xi_i \geq 0,
\]

\noindent where $C > 0$ is a regularization parameter that controls the trade-off between maximizing the margin and minimizing the classification error over all $N$  examples, and $\phi(\cdot)$ is a feature mapping induced by a kernel function $K(\mathbf{x}_i, \mathbf{x}_j) = \langle \phi(\mathbf{x}_i), \phi(\mathbf{x}_j) \rangle$. In our implementation, we use the radial basis function (RBF) kernel:
\[
K(\mathbf{x}, \mathbf{x}') = \exp\left(-\gamma \|\mathbf{x} - \mathbf{x}'\|^2\right),
\]
where $\gamma$ is a kernel width parameter tuned via cross-validation.

\subsubsection{Random Forest (RF)}

RF is an ensemble learning method that builds a collection of decision trees, each trained on a random subset of the training data and feature set. The final prediction is obtained through majority voting in classification tasks. The strength of Random Forest lies in its ability to reduce variance while maintaining low bias, making it robust against overfitting~\cite{breiman2001random}.

Formally, the prediction $\hat{y}$ for an input $\mathbf{x}$ is given by:

\begin{equation}
\hat{y} = \text{mode}\left\{ h_m(\mathbf{x}) \right\}_{m=1}^{M}
\end{equation}

\noindent where $h_m(\cdot)$ is the $m$-th decision tree in the ensemble, and $M$ is the total number of trees. Each tree is trained on a random sample drawn with replacement from the training data, and at each split, a random subset of features is considered to introduce decorrelation among trees.

\subsubsection{eXtreme Gradient Boosting (XGB)}

XGB (eXtreme Gradient Boosting) is a highly optimized and scalable implementation of gradient boosting machines, specifically designed for superior performance on structured tabular data~\cite{chen2015xgboost}. The algorithm constructs an ensemble of weak learners—typically decision trees—in a sequential manner. At each iteration, a new tree is trained to minimize the residual errors of the current ensemble, thereby progressively refining the model's predictive accuracy.

We employ a gradient boosting framework where the prediction for an input instance $\mathbf{x}_i$ at a given iteration $q$ is formulated as an additive sum of $q$ individual decision trees. This prediction, denoted as $\widehat{y}_i^{(q)}$, is given by:

\begin{equation}
\widehat{y}_i^{(q)} = \sum_{k=1}^{q} f_k(\mathbf{x}_i), \quad f_k \in \mathcal{F} \, .
\end{equation}

Here, $\mathcal{F}$ represents the function space of regression trees, and each $f_k$ signifies a single decision tree.

The model undergoes iterative optimization by minimizing a regularized objective function, $\mathcal{L}^{(q)}$, at each boosting step. This objective is defined as:

\begin{equation}
\mathcal{L}^{(q)} = \sum_{i=1}^{N} \ell(y_i, \hat{y}_i^{(q)}) + \sum_{k=1}^{q} \Omega(f_k)
\end{equation}

The objective function $\mathcal{L}^{(q)}$ comprises two distinct components:
\begin{enumerate}
    \item A differentiable \textbf{loss function}, $\ell(y_i, \hat{y}_i^{(q)})$, which quantifies the discrepancy between the true label $y_i$ and the current predicted value $\hat{y}_i^{(q)}$. During optimization, the derivatives of this loss function are computed with respect to the model's current predictions, specifically $\hat{y}_i^{(q-1)}$ from the previous iteration. For regression tasks, squared error is a common choice for $\ell$. For multi-class classification, \textbf{Categorical Cross-Entropy Loss} is typically employed.
    \item A \textbf{regularization term}, $\Omega(f_k)$, which penalizes the complexity of the model to prevent overfitting.
\end{enumerate}

The regularization term $\Omega(f)$ for an individual tree $f$ is explicitly defined as:

\begin{equation}
\Omega(f) = \gamma Q + \frac{1}{2} \lambda \sum_{j=1}^{Q} w_j^2
\end{equation}

In this definition, $Q$ denotes the number of leaves in the tree, and $w_j$ is the weight (output value) assigned to the $j$-th leaf. The hyperparameter $\gamma$ introduces a penalty for each additional leaf node, while $\lambda$ serves as the L2 regularization coefficient applied to the leaf weights. This regularization scheme is crucial for controlling model complexity and enhancing its generalization ability to unseen data.

To facilitate this iterative optimization process, the model's prediction at iteration $q$ can also be expressed recursively:

\begin{equation}
\hat{y}_i^{(q)} = \hat{y}_i^{(q-1)} + f_q(\mathbf{x}_i)
\end{equation}

Here, $\hat{y}_i^{(q-1)}$ represents the aggregate prediction accumulated from the preceding $q-1$ trees. The term $f_q(\mathbf{x}_i)$ is the prediction contributed by the newly added tree at the current iteration $q$, which is specifically trained to approximate the negative gradient (often referred to as pseudo-residual) of the loss function with respect to $\hat{y}_i^{(q-1)}$.

XGB's strengths lie in its high computational efficiency, built-in regularization, and scalability to large datasets. These qualities make it particularly well-suited for learning from microstate features in tabular form.

\subsection{Explainability with SHAP}
\label{subsec:shap}

To interpret the contribution of individual input features to the model's predictions, we employed SHAP (SHapley Additive exPlanations) \cite{lundberg2017shap}, a unified framework grounded in cooperative game theory. SHAP values offer a theoretically consistent and locally accurate measure of feature importance, applicable across a wide range of models including tree ensembles (XGB or RFs) and kernel-based models (SVM).

\subsubsection{Shapley Value Foundation.}
The SHAP framework is based on the concept of Shapley values, originally developed in the context of cooperative games. Consider a model $f: \mathbb{R}^d \rightarrow \mathbb{R}$ and a prediction instance $x = (x_1, \dots, x_d)$. The goal is to express the model output $f(x)$ as a sum of contributions from each feature:
\begin{equation}
    f(x) = \phi_0 + \sum_{i=1}^d \phi_i,
\end{equation}
where $\phi_0 = \mathbb{E}_{\mathbf{x}}[f(\mathbf{x})]$ is the expected model output under the data distribution and $\phi_i$ denotes the contribution of feature $i$ to the deviation from this baseline.

The value $\phi_i$ is defined via the Shapley value:
\begin{equation}
    \phi_i = \sum_{S \subseteq N \setminus \{i\}} \frac{|S|! (d - |S| - 1)!}{d!} \left[ f_{S \cup \{i\}}(x_{S \cup \{i\}}) - f_S(x_S) \right],
    \label{eq:shap_exact}
\end{equation}
where $N = \{1, 2, \dots, d\}$ is the set of all feature indices, $S$ is a subset of features excluding $i$, and $f_S(x_S)$ denotes the expected output of the model when only the features in $S$ are known:
\begin{equation}
    f_S(x_S) = \mathbb{E}_{\mathbf{x}_{\bar{S}}}[f(x_S, \mathbf{x}_{\bar{S}})],
\end{equation}
with $\bar{S}$ being the complement of $S$.

\subsubsection{Computational Efficiency via TreeSHAP.}
For RFs and XGB, we utilize the TreeSHAP algorithm \cite{lundberg2020local}, which enables exact computation of SHAP values in polynomial time. TreeSHAP leverages the tree structure to recursively compute conditional expectations, achieving a runtime complexity of $\mathcal{O}(TLD^2)$, where $T$ is the number of trees, $L$ is the maximum number of leaves per tree, and $D$ is the maximum tree depth.

\subsubsection{SHAP for Non-Tree Models.}
For non-tree models such as SVM, where exact SHAP computation is intractable, we employ the KernelSHAP method. This approach approximates the Shapley values via a weighted linear regression on samples from the power set of features, providing a model-agnostic estimation of $\phi_i$ under the additive feature attribution framework.

\subsubsection{SHAP Axioms and Interpretability.}
SHAP values satisfy key axioms that ensure reliable interpretability:
\begin{itemize}
    \item \textbf{Local Accuracy (Efficiency):} The attributions sum to the prediction difference.
    \item \textbf{Missingness:} Features not in the model receive zero attribution.
    \item \textbf{Consistency:} If a model changes so that the marginal contribution of a feature increases, its SHAP value does not decrease.
\end{itemize}

By leveraging SHAP values, we obtain a consistent and model-agnostic explanation of the influence of individual features across our ensemble and kernel-based predictive frameworks, enhancing the transparency and trustworthiness of our models.

\section{Experimental Setup} \label{exp}

To assess the performance of our proposed microstate-based classification framework, we conducted a comprehensive set of experiments across two EEG datasets: the CAUEEG dataset and the General Hospital of Thessaloniki dataset. The dataset was partitioned into training and testing sets, ensuring that subject-level separation was maintained to prevent information leakage. We evaluated three traditional ML classifiers, namely SVM, RF, and XGB,using microstate-derived features, and employed SHAP-based analysis for post hoc interpretability. Hyperparameters for each model were optimized via grid search with cross-validation on the training data.

\subsection{Evaluation Metrics}
\label{subsec:evaluation_metrics}

To quantitatively assess the performance of our multi-class classification models, we employed a suite of standard evaluation metrics: accuracy, precision, recall, and F1-score. These metrics provide complementary views on model performance, including accuracy, class-wise discrimination, and robustness to imbalanced data
distributions.
\begin{itemize}

\item \textbf{Accuracy.}
Accuracy represents the ratio of correctly predicted instances to the total number of samples:
\begin{equation}
    \text{Accuracy} = \frac{1}{n} \sum_{i=1}^n \mathbb{I}(\hat{y}_i = y_i),
\end{equation}
where $n$ is the number of instances, $y_i$ is the true class label, $\hat{y}_i$ is the predicted label, and $\mathbb{I}(\cdot)$ is the indicator function.

\item \textbf{Precision, Recall, and F1-Score.}
For each class $c \in \mathcal{C}$, we define:
   \begin{itemize}
    \item \textbf{Precision} (Positive Predictive Value) quantifies the proportion of true positives among all predicted positives for class $c$:
    \begin{equation}
        \text{Precision}_c = \frac{\text{TP}_c}{\text{TP}_c + \text{FP}_c},
    \end{equation}
    where $\text{TP}_c$ and $\text{FP}_c$ denote the number of true positives and false positives, respectively.

    \item \textbf{Recall} (Sensitivity or True Positive Rate) measures the proportion of true positives correctly identified among all actual positives:
    \begin{equation}
        \text{Recall}_c = \frac{\text{TP}_c}{\text{TP}_c + \text{FN}_c},
    \end{equation}
    where $\text{FN}_c$ is the number of false negatives for class $c$.

    \item \textbf{F1-Score} is the harmonic mean of precision and recall, providing a balance between the two:
    \begin{equation}
        \text{F1}_c = \frac{2 \cdot \text{Precision}_c \cdot \text{Recall}_c}{\text{Precision}_c + \text{Recall}_c}.
    \end{equation}
    
    \end{itemize}

\item \textbf{Macro-Averaging.}
Given the multi-class nature of our problem, we adopted macro-averaging to aggregate the per-class metrics. This approach computes the unweighted mean across all classes:
\begin{equation}
    \text{Macro-Precision} = \frac{1}{|\mathcal{C}|} \sum_{c \in \mathcal{C}} \text{Precision}_c,
\end{equation}
with analogous formulations for macro-recall and macro-F1. This averaging strategy ensures that each class contributes equally to the overall performance, regardless of its frequency in the dataset.
\end{itemize}

\subsection{Implementation Details}
We performed inter-subject multi-class classification. Implementation details of our microstate analysis framework on the CAUEEG and the General Hospital of Thessaloniki dataset are as follows:
 EEG signals are preprocessed using the standardized procedures described in Subsection~\ref{preproc}. The EEG-MSAF offers a configurable interface that enables users to select specific frequency bands of interest (e.g., alpha (8-12 Hz), beta (12-30 Hz)) before microstate segmentation, allowing for frequency-resolved analysis of brain network dynamics. 

An interactive module within the framework supports microstate identification and visual inspection. This interface allows users to visualize and label canonical microstate maps (A, B, C, F). The microstate features are extracted for each group and then saved in a structured, tabular format for downstream analysis. 

Using the extracted features, we implement three versions of our proposed EEG-MSAF framework by varying the parameters among the three final classifiers: SVM, RFs and XGB. To distinguish between them, we refer to these variants as \textbf{EEG-MSAF-SVM}, \textbf{EEG-MSAF-RF}, and \textbf{EEG-MSAF-XGB}, respectively.

For the Random Forest-based classifier, we performed a grid search over the number of estimators \{100, 200, 300\}, maximum tree depth \{5, 10, 15\}, and minimum samples per split \{2, 4\}, using 5-fold cross-validation. SVM-based models were trained using the radial basis function (RBF) kernel, with hyperparameters $C \in \{0.1, 1, 10, 100\}$ and $\gamma \in \{0.0001, 0.001, 0.05\}$, and the one-vs-rest strategy was employed for multi-class classification. For XGB-based models, we tuned the number of boosting rounds \{100, 200\}, learning rate \{0.0001, 0.001, 0.05\}, and maximum depth \{3, 6, 10\}, using early stopping with a patience of 10 rounds to mitigate overfitting.

All experiments were conducted on a workstation equipped with an Intel Core i7 CPU and 32 GB of RAM. The implementation was developed in Python 3.9, leveraging standard libraries including \texttt{mne}, \texttt{pycrostates}, \texttt{scikit-learn}, \texttt{xgboost}, and \texttt{shap}. To enhance interpretability, the framework integrates SHAP (SHapley Additive exPlanations), which computes post hoc feature attributions for each classifier. Class-specific SHAP value analysis is performed to rank the importance of each microstate feature in distinguishing between NC, MCI, and DEM classes.

\subsection{Baseline Models}
To benchmark the performance of our proposed EEG-MSAF framework, we compared it against state-of-the-art baseline models. These include CeedNet~\cite{kim2023deep} for the CAUEEG dataset and EEGConvNeXt~\cite{acharya2025eegconvnext} for the dataset from General Hospital of Thessaloniki, which leverage DL architectures tailored for EEG signal classification. CeedNet is trained on EEG signals, while EEGConvNeXt adopts the ConvNeXt architecture, known for its strong performance in computer vision, to address the specific characteristics of EEG signals.

To ensure a fair comparison, our framework was trained on the same dataset used by the baseline models.

\section{Results}\label{results}

\begin{figure*}[htbp]
\centering

\begin{minipage}[b]{0.50\textwidth}
    \centering
    \includegraphics[width=\linewidth,trim={0cm 0cm 0cm 0cm},clip]{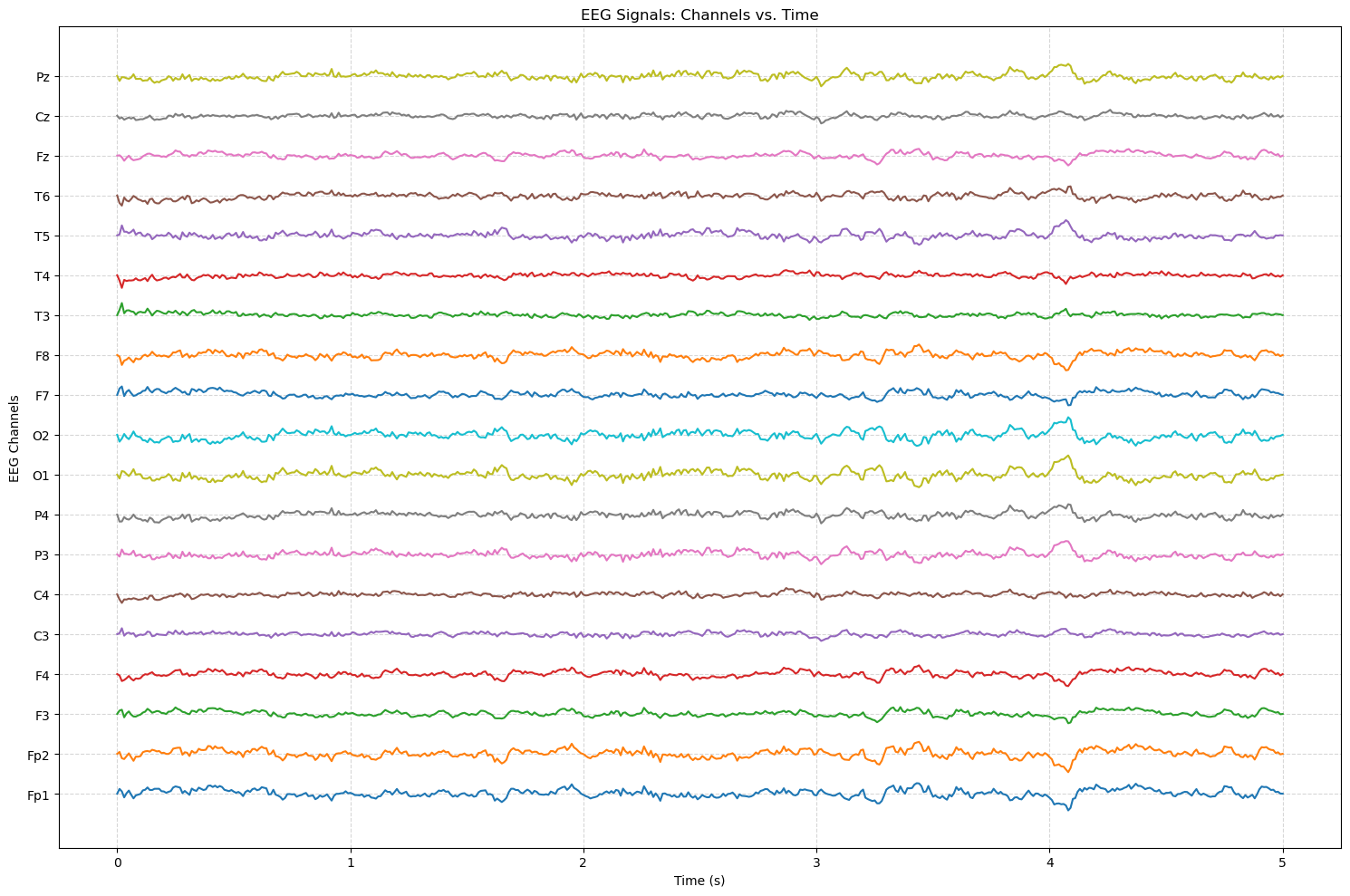}
    \\ (i)\medskip

    \includegraphics[width=1.10\linewidth,trim={0.1cm 0cm 0cm 0cm},clip]{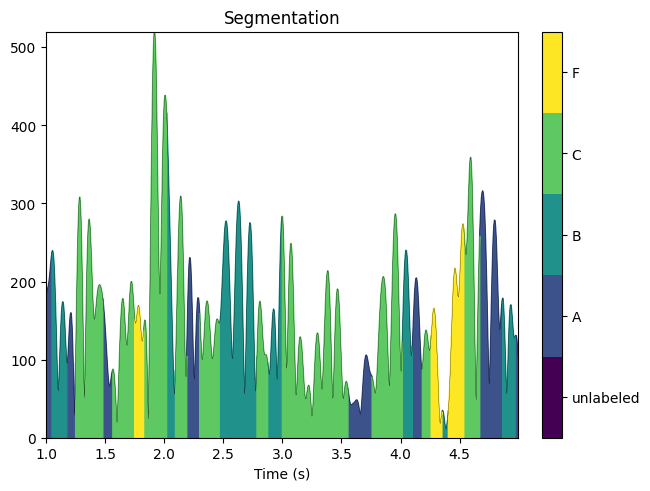}
    \\ (ii)
\end{minipage}
\hfill
\begin{minipage}[t]{0.48\textwidth}
    \centering
    \vspace{-12cm} 

    \includegraphics[width=\linewidth, trim={0cm 18cm 0cm 5cm},clip]{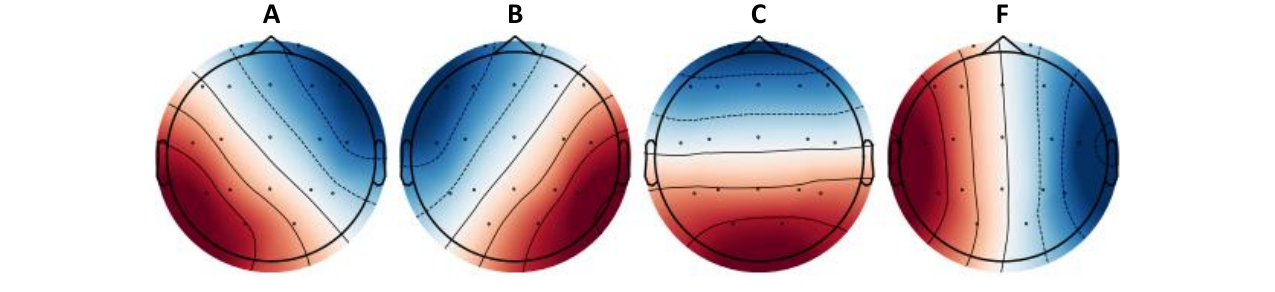}
    \\ (iii)\medskip

    \includegraphics[width=\linewidth,trim={0cm 18cm 0cm 1cm},clip]{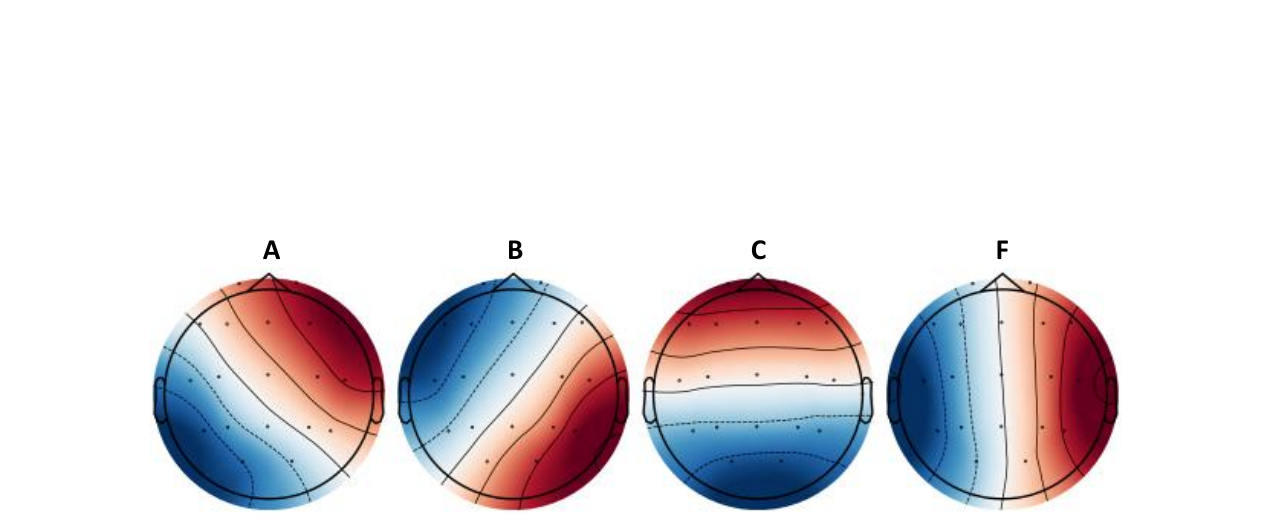}
    \\ (iv)\medskip

    \includegraphics[width=\linewidth,trim={0cm 18cm 0cm 1cm},clip]{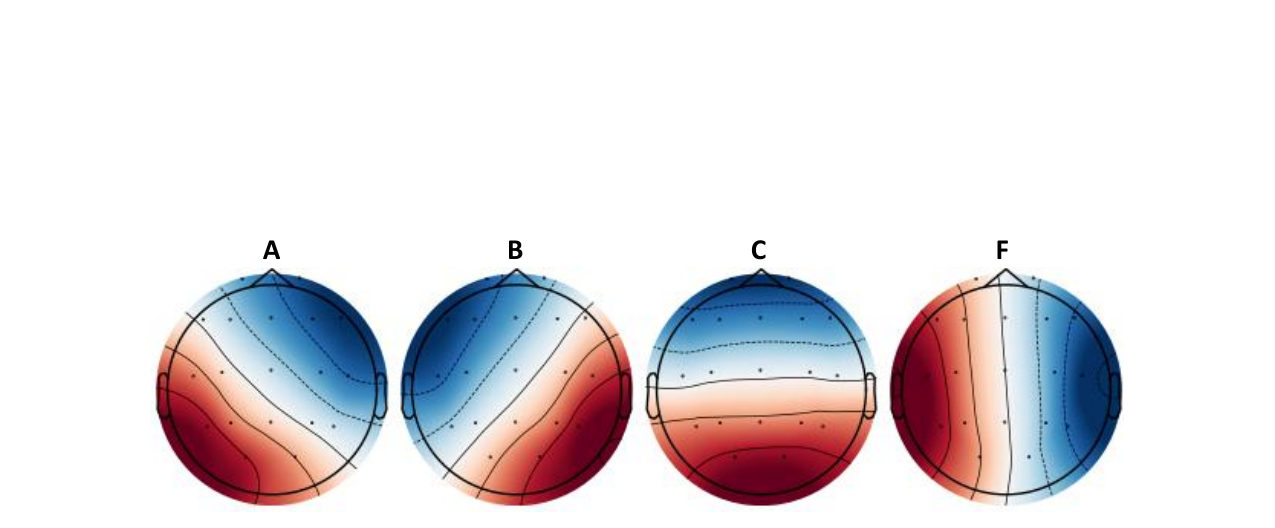}
    \\ (v)
\end{minipage}

\caption{
\textbf{Representative EEG signal and corresponding microstate segmentation with group-specific topographies.}
(i) A 5-second segment of eyes-closed resting-state EEG is shown. 
(ii) The same EEG trace is segmented into a sequence of canonical microstate classes (A, B, C, F), where each time point is color-coded according to its assigned microstate. The vertical height of the color bands represents the instantaneous Global Field Power (GFP), reflecting the amplitude of the EEG field at each moment. This segmentation reveals both the temporal dynamics and stability of the underlying brain states.
(iii–v) Normalized group-averaged scalp topographies of the four canonical microstate classes (A, B, C, F) for three diagnostic groups: (iii) NC group, (iv) MCI group, and (v) DEM group. Each topography represents the mean spatial voltage distribution across epochs assigned to a given microstate class, averaged across all subjects within the respective group. Areas of opposite polarity are depicted in red and blue. The nose is oriented upward, and the left ear is to the left. These topographies capture both the preserved and altered spatial features of microstate patterns across clinical groups.
}

\label{fig:microstate_topographies}
\end{figure*}

Figure~\ref{fig:microstate_topographies} displays the canonical topographies of the four extracted microstate classes (A, B, C, F), derived from the CAUEEG dataset. These spatial patterns closely resemble those originally reported by ~\cite{koenig2002millisecond, tarailis2024functional,brechet2019capturing}, confirming the neurophysiological plausibility of our microstate segmentation. Specifically, microstate A exhibits a left occipital to right frontal orientation, while microstate B presents a mirrored pattern from the right occipital to left frontal regions. Microstate C demonstrates a symmetric occipital to prefrontal distribution, and microstate F shows a left-lateralized configuration.

\begin{figure*}[ht]
    \centering
    \begin{minipage}{0.49\textwidth}
        \centering       \includegraphics[width=\textwidth, trim={0.0cm 0cm 0.3cm 0.2cm}, clip]{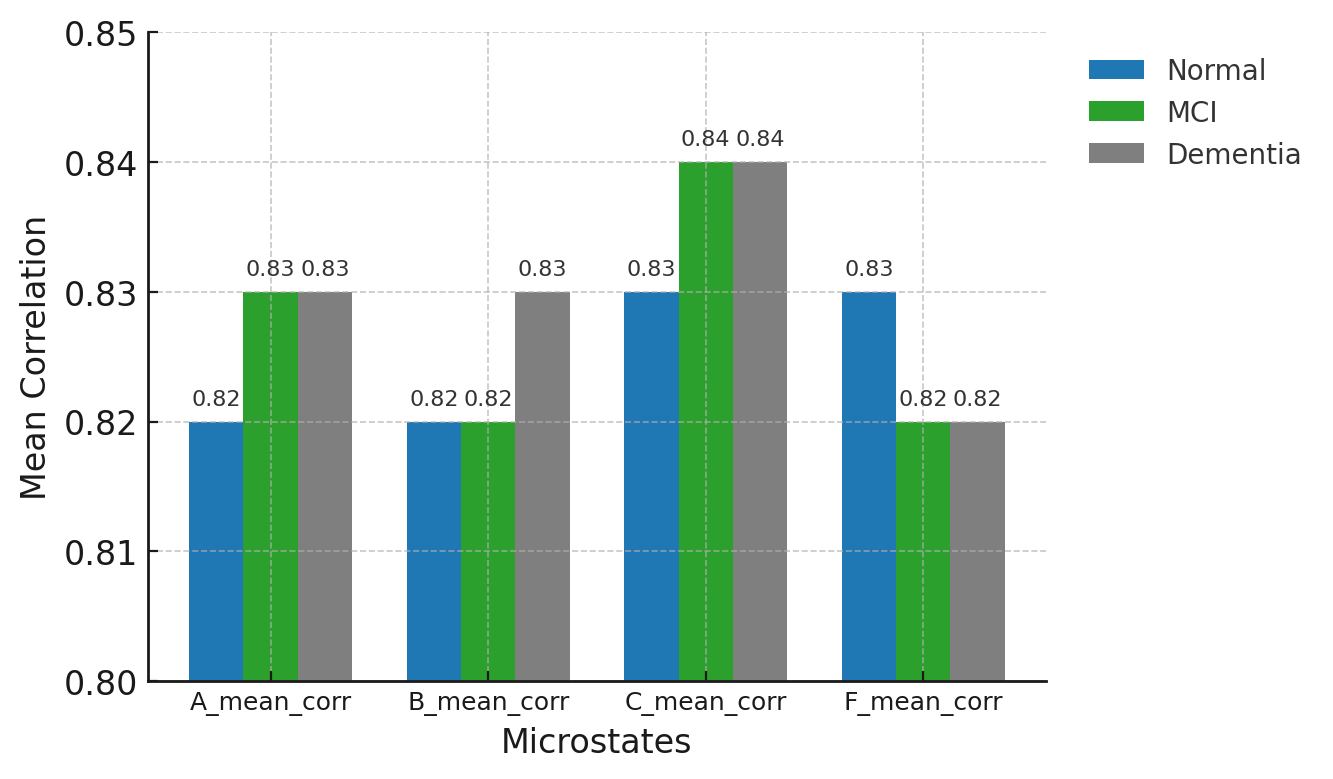}
    \end{minipage}
    \hfill
    \begin{minipage}{0.49\textwidth}
       \centering
 \includegraphics[width=\textwidth, trim={0.0cm 0cm 0.3cm 0.5cm}, clip]{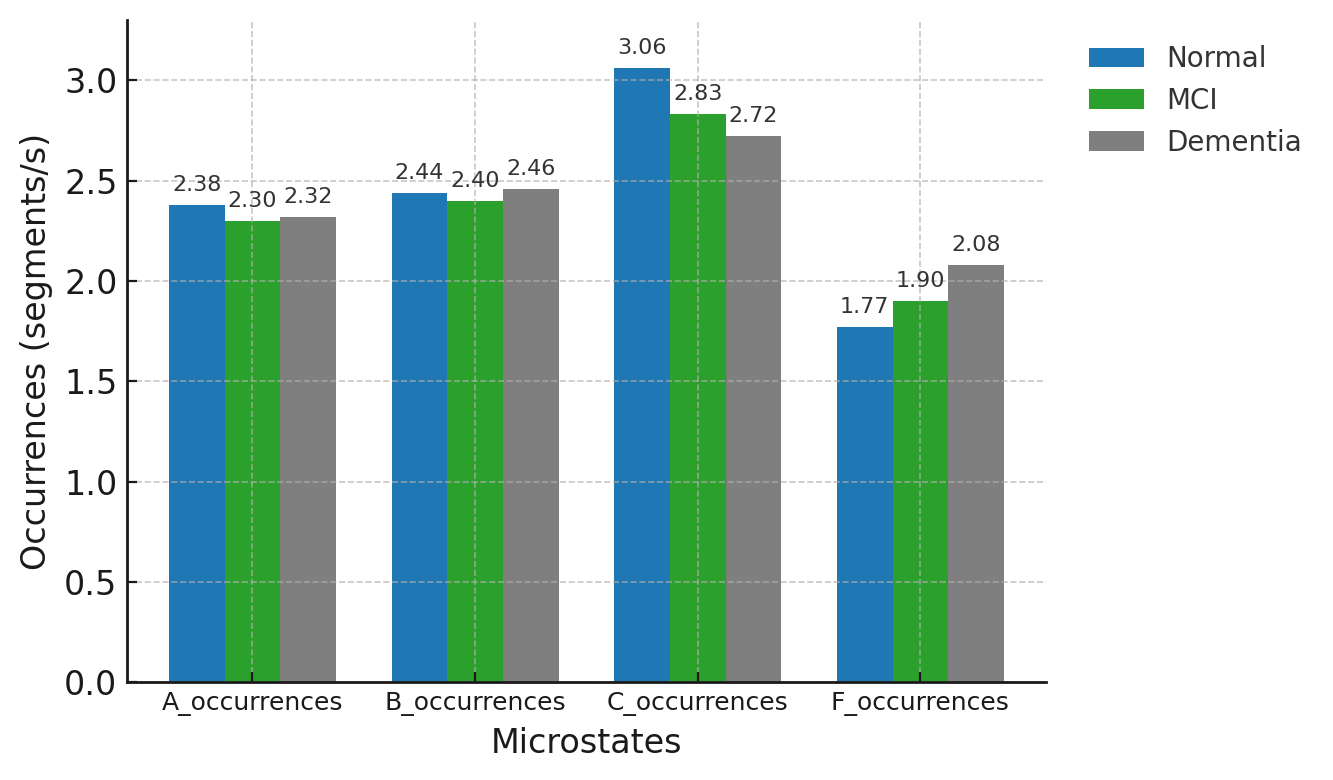}
    \end{minipage}
    \caption{\protect
    \textbf{(Left)} Mean correlation and 
    \textbf{(Right)} Mean Occurrences of EEG microstates (A, B, C, F) associated with the theta frequency band (4-8 Hz) across NC, MCI, and DEM groups.
    These two features provide a complementary view of spatiotemporal brain activity. Notably, microstate C shows a consistent decline in occurrence with decreasing progression. In contrast, microstates  B and F exhibit increased occurrence in DEM group.}\label{fig:microstate_features}
\end{figure*}

\begin{figure}[htb!]
    \centering
    \includegraphics[width=0.47\textwidth ,height=0.30\textwidth ,trim={0.2cm 0cm 0cm 0cm},clip]{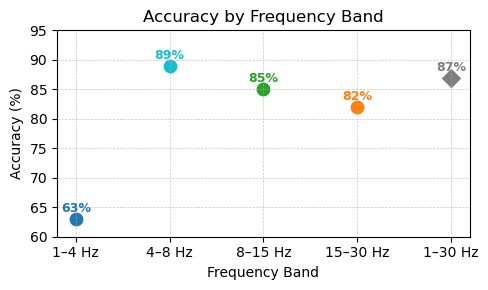}
   
   \caption{Classification accuracy (NC, MCI, DEM) across EEG frequency bands.Illustrating the EEG-MSAF-SVM model´s performance across different frequency bands on the CAUEEG dataset. The 4–8 Hz (theta) band achieves the highest classification accuracy.}
   \label{fig:bands_vs_acc}
\end{figure}

To evaluate the classification performance of our end-to-end explainable framework, we tested three traditional ML models: SVM, RFs,  and XGB. We conducted experiments on two publicly available datasets: the CAUEEG dataset and the Thessaloniki Hospital dataset. 
As presented in Table~\ref{tab:classification_performance}, the EEG-MSAF-SVM model achieved the highest classification performance, attaining an accuracy of $0.89\pm{0.01}$ under 5-fold cross-validation. Furthermore, we also conducted experiments across distinct EEG bands. Notably, the theta band (4–8 Hz) yielded the best performance, as illustrated in Figure~\ref{fig:bands_vs_acc}. We further tested our framework on the second dataset from Thessaloniki hospital, which involved classification among NC, Frontotemporal Dementia (FTD), and AD groups. EEG-MSAF-SVM again achieved the highest accuracy of $0.95\pm{0.01}$, outperforming EEGConvNeXt, a DL baseline (see Table~\ref{tab:cf_greek}). The results highlight the effectiveness and robustness of our proposed framework, based on traditional ML models, particularly EEG-MSAF-SVM, in capturing discriminative EEG patterns relevant for DEM diagnosis.

\begin{figure}[htb!]
    \centering

    \includegraphics[width=0.47\textwidth,height=5.6 cm, trim={0.5cm 0.5cm 0cm 0cm},clip]{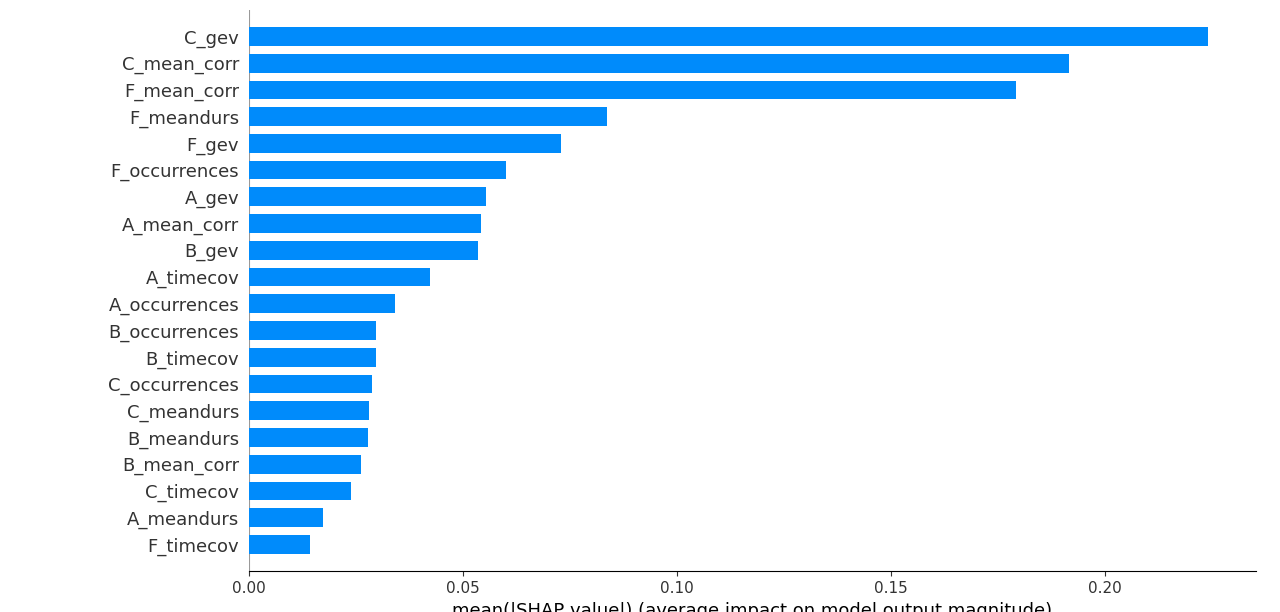}
    \label{fig:normal}
    \par\smallskip
    \centering (a)

    \medskip

    \includegraphics[width=0.45\textwidth,height=5.6 cm,trim={0.5cm 0.5cm 0cm 0cm},clip]{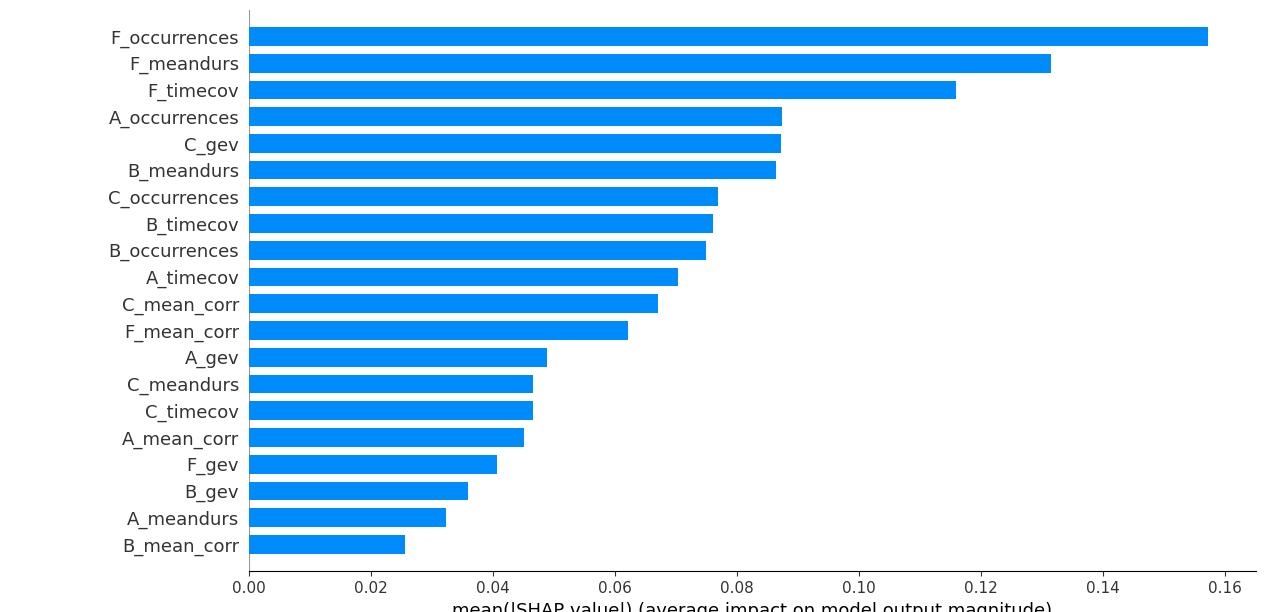}
    \label{fig:mci}
    \par\smallskip
    \centering (b)

    \medskip

    \includegraphics[width=0.45\textwidth,height=5.6 cm,trim={0.5cm 0.5cm 0cm 0cm},clip]{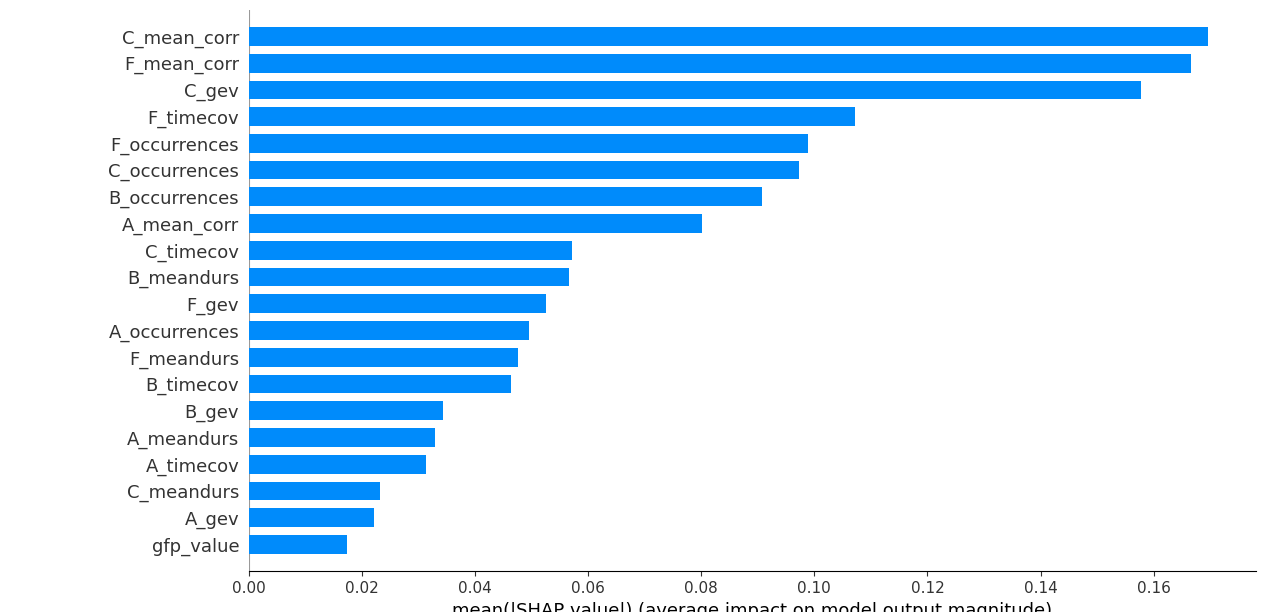}
    \par\smallskip
    \label{fig:dementia}
    \centering (c)

    \caption{SHAP-based feature importance rankings from the best-performing SVM model across the three groups in the CAUEEG dataset: (a) NC, (b) MCI, and (c) DEM. Each plot shows the contribution of individual EEG microstate features to the model´s predictions. The SHAP scores consistently attribute high importance to microstate correlation and occurrence features, underscoring their relevance in distinguishing between cognitive states.}

    \label{fig:shap}
\end{figure}

We employed SHAP (SHapley Additive exPlanations) to interpret the decision processes of the trained SVM models. Separate SHAP analyses were conducted for each clinical group to identify which microstate features contributed most significantly to classification. As shown in Figure~\ref{fig:shap}, mean correlation and occurrence metrics were consistently ranked as top contributors across all groups. Notably, microstate C's mean correlation and duration were among the most important features in the DEM group, aligning with previous studies reporting altered temporal properties in this state~\cite{musaeus2019microstates}. The alignment of model-derived feature rankings with known neurophysiological patterns supports the interpretability and reliability of our method.

To contextualize the performance of our interpretable framework, we compared its accuracy with that of state-of-the-art DL models. On the CAUEEG dataset, CEEDNET~\cite{kim2023deep}, a DL model, served as the baseline, while on the Thessaloniki Hospital dataset, EEGConvNeXt~\cite{acharya2025eegconvnext} was taken as the baseline model. We achieved a notable 19.3\% improvement with our method, particularly with the EEG-MSAF-SVM model, compared to CEEDNET.  Although these deep networks achieved competitive accuracy, they lack interpretability. In contrast, our EEG-MSAF-SVM model not only outperformed these baselines but also provided transparent decision-making through feature attribution. This emphasizes the value of our approach in clinical EEG analysis, offering both high accuracy and interpretability. Additionally, all our models are lightweight compared to DL baselines.

\begin{table}[htb]
\definecolor{headercolor}{RGB}{224, 217, 202} 
\definecolor{rowgray}{gray}{0.95}             
\centering
\caption{Classification performance of different models on the CAUEEG dataset. The SVM model achieved the best balance across metrics.}
\label{tab:classification_performance}
\resizebox{\columnwidth}{!}{%
\begin{tabular}{lcccc}
\toprule
\rowcolor{headercolor}
\textbf{Model} \rule{0pt}{2.6ex} & \textbf{Accuracy} & \textbf{Precision} & \textbf{Recall} & \textbf{F1-score} \\
\midrule
CeedNet~\cite{kim2023deep}              & 0.746           & 0.743             & 0.726           & 0.730             \\
EEG-MSAF-Random Forest                  & 0.65$\pm$0.01   & 0.66$\pm$0.01     & 0.65$\pm$0.01   & 0.65$\pm$0.01     \\
EEG-MSAF-XGB                        & 0.78$\pm$0.02   & 0.77$\pm$0.00     & 0.77$\pm$0.01   & 0.77$\pm$0.01     \\
\rowcolor{gray!15}
EEG-MSAF-SVM                            & \textbf{0.89$\pm$0.01} & \textbf{0.88$\pm$0.01} & \textbf{0.89$\pm$0.01} & \textbf{0.88$\pm$0.01} \\
\bottomrule
\end{tabular}
}
\end{table}


\begin{table}[htb]
\definecolor{headercolor}{RGB}{224, 217, 202}
\definecolor{rowgray}{gray}{0.95}
\centering
\caption{Classification performance of different models on the dataset from the General Hospital of Thessaloniki. The EEG-MSAF-SVM model achieved the best balance across metrics.}
\label{tab:cf_greek}
\resizebox{\columnwidth}{!}{%
\begin{tabular}{lcccc}
\toprule
\rowcolor{headercolor}
\textbf{Model} \rule{0pt}{2.6ex} & \textbf{Accuracy} & \textbf{Precision} & \textbf{Recall} & \textbf{F1-score} \\
\midrule
EEGConvNeXt~\cite{acharya2025eegconvnext}         &   0.9570           & 0.9608             & 0.9566           & 0.9587            \\
EEG-MSAF-Random Forest  & 0.86$\pm{0.01}$         & 0.88 $\pm{0.01}$              & 0.88 $\pm{0.01}$           & 0.86$\pm{0.01}$            \\
EEG-MSAF-XGB        & 0.73$\pm{0.02}$           & 0.71$\pm{0.02}$               & 0.71$\pm{0.02}$            & 0.71$\pm{0.02}$            \\
\rowcolor{gray!15}
EEG-MSAF-SVM            & \textbf{0.95}$\pm{0.01}$    & \textbf{0.96}$\pm{0.01}$      & \textbf{0.96}$\pm{0.01}$   & \textbf{0.96}$\pm{0.01}$  \\
\bottomrule
\end{tabular}
}
\end{table}



\begin{table}[ht]
\definecolor{headercolor}{RGB}{224, 217, 202} 
\definecolor{rowgray}{gray}{0.95}             
\centering
\caption{Results of statistical tests comparing EEG microstate features across NC, MCI, and DEM groups.}
\label{tab:stat_tests}
\begin{tabular}{|l|l|c|c|}
\hline
\rowcolor{headercolor}
\textbf{Feature} & \textbf{Test} & \textbf{Statistic} & \textbf{p-value}  \\
\hline

\rowcolor{white}
A\_mean\_corr      & Kruskal–Wallis &67.73 & $<0.0001$   \\
\rowcolor{rowgray}
B\_mean\_corr      & Kruskal–Wallis & 51.7 & $<0.0001$   \\
\rowcolor{white}
C\_mean\_corr      & Kruskal–Wallis & 81.75  & $<0.0001$  \\
\rowcolor{rowgray}
F\_mean\_corr      & Kruskal–Wallis & 5.08  & $0.079$   \\
\rowcolor{white}
A\_occurrences     & Kruskal–Wallis & 21.58 & $<0.0001$   \\
\rowcolor{rowgray}
B\_occurrences     & Kruskal–Wallis & 8.94 & $<0.0001$   \\
\rowcolor{white}
C\_occurrences     & Kruskal–Wallis & 180.04  & $<0.0001$   \\
\rowcolor{rowgray}
F\_occurrences     & Kruskal–Wallis & 114.96 & $<0.0001$  \\
\rowcolor{white}

\hline
\end{tabular}
\end{table}
\begin{table}[ht]
\definecolor{headercolor}{RGB}{224, 217, 202} 
\definecolor{rowgray}{gray}{0.95}             
\centering
\caption{Post-hoc comparison results for EEG microstate features. Adjusted p-values are computed using Bonferroni correction for Kruskal–Wallis tests and Tukey's HSD for ANOVA. Comparisons with $p < 0.05$ are considered statistically significant.}
\label{tab:posthoc_results}
\resizebox{\columnwidth}{!}{%
\begin{tabular}{|l|l|l|c|}
\hline
\rowcolor{headercolor}
\textbf{Feature} & \textbf{Test} & \textbf{Comparison} & \textbf{Adjusted p-value} \\
\hline
\rowcolor{white}
A\_mean\_corr  & Kruskal–Wallis & DEM vs MCI      & $4.45\times10^{-10}$ \\
\rowcolor{rowgray}
B\_mean\_corr  & Kruskal–Wallis & DEM vs MCI      & $7.54\times10^{-07}$ \\
\rowcolor{white}
C\_mean\_corr  & Kruskal–Wallis & DEM vs MCI      & $3.14\times10^{-06}$ \\
\rowcolor{rowgray}
A\_occurrences & Kruskal–Wallis & DEM vs MCI      & $1.00$                \\
\rowcolor{white}
B\_occurrences & Kruskal–Wallis & DEM vs MCI      & $1.34\times10^{-02}$  \\
\rowcolor{rowgray}
C\_occurrences & Kruskal–Wallis & DEM vs MCI      & $8.40\times10^{-04}$  \\
\rowcolor{white}
F\_occurrences & Kruskal–Wallis & DEMa vs MCI      & $1.16\times10^{-07}$  \\
\rowcolor{rowgray}
A\_mean\_corr  & Kruskal–Wallis & DEM vs NC   & $6.35\times10^{-15}$ \\
\rowcolor{white}
B\_mean\_corr  & Kruskal–Wallis & DEM vs NC   & $4.22\times10^{-12}$ \\
\rowcolor{rowgray}
C\_mean\_corr  & Kruskal–Wallis & DEM vs NC   & $5.21\times10^{-19}$ \\
\rowcolor{white}
A\_occurrences & Kruskal–Wallis & DEM vs NC   & $2.22\times10^{-03}$ \\
\rowcolor{rowgray}
B\_occurrences & Kruskal–Wallis & DEM vs NC   & $1.00$               \\
\rowcolor{white}
C\_occurrences & Kruskal–Wallis & DEM vs NC   & $5.81\times10^{-36}$ \\
\rowcolor{rowgray}
F\_occurrences & Kruskal–Wallis & DEM vs NC   & $3.68\times10^{-26}$ \\
\rowcolor{rowgray}
A\_mean\_corr  & Kruskal–Wallis & MCI vs NC        & $0.391$              \\
\rowcolor{white}
B\_mean\_corr  & Kruskal–Wallis & MCI vs NC        & $0.145$              \\
\rowcolor{rowgray}
C\_mean\_corr  & Kruskal–Wallis & MCI vs NC        & $3.42\times10^{-05}$ \\
\rowcolor{white}
A\_occurrences & Kruskal–Wallis & MCI vs NC        & $4.11\times10^{-05}$ \\
\rowcolor{rowgray}
B\_occurrences & Kruskal–Wallis & MCI vs NC        & $9.15\times10^{-02}$ \\
\rowcolor{white}
C\_occurrences & Kruskal–Wallis & MCI vs NC        & $1.41\times10^{-21}$ \\
\rowcolor{rowgray}
F\_occurrences & Kruskal–Wallis & MCI vs NC        & $1.12\times10^{-07}$ \\
\hline
\end{tabular}
}
\end{table}


\section{Discussion}\label{discuss}

 We showed the analysis of EEG microstate dynamics across NC, MCI, and DEM groups using both statistical measures (mean correlation and occurrence) and model-derived explanations (SHAP-based feature importance). Our findings offer critical insights into the neurophysiological changes associated with cognitive decline and underscore the diagnostic utility of interpretable features derived from EEG microstates. 
 In this study, we propose an end-to-end explainable framework for EEG-based DEM classification. Leveraging microstate-derived features and a traditional ML model, namely SVM, our approach achieves state-of-the-art performance. Beyond classification accuracy, we provide a detailed analysis of EEG microstate dynamics across NC, MCI, and DEM groups using both statistical descriptors—mean correlation and occurrence, and model-driven explanations via SHAP based feature importance. Our results provide critical insight into the underlying neurophysiological changes associated with cognitive decline and highlight the diagnostic value of interpretable microstate features in understanding and differentiating stages of DEM.

\subsection{Opposing trajectories of microstates C and F}

In Figure~\ref{fig:microstate_features}, we observe that microstate \textbf{C}—functionally linked to the DMN and medial temporal structures, consistently shows a reduction in occurrence and only a marginal rise in spatial coherence, whereas the anterior DMN-related microstate F exhibits the opposite pattern, higher occurrence but declining coherence from normal aging through MCI to DEM. This "pull and push" pattern aligns with the large-scale network degeneration/imbalance hypothesis observed with fMRI in AD~\cite{seeley2009neurodegenerative}. This pattern also reflects a breakdown in salience network functionality, aligning with previous literature that associates microstate C with cognitive control, object-visual thinking, attention reorientation and decision-making processes~\cite{nishida2013eeg, khanna2015microstates, musaeus2019microstates}. By contrast, microstates \textbf{B} and \textbf{F} exhibit higher occurrence rates in the DEM group, despite relatively stable or non-monotonic correlation levels. This divergence suggests a shift toward more frequent but potentially less stable brain state transitions, possibly reflecting either a compensatory mechanism or network dysregulation~\cite{dierks1997eeg}.

\subsection{Complementarity of Correlation and Occurrence}

Observing the feature rankings in Figure~\ref{fig:shap} reveals that mean correlation and occurrence are consistently ranked as the top features in all the groups (NC, MCI, and DEM). Thus, we conducted a complementary analysis between the mean correlation and the occurrence features. While correlation reflects the internal coherence of a given microstate, occurrence captures its engagement frequency. Notably, features such as  \texttt{F\_mean\_corr} and \texttt{F\_occurrences} demonstrate high importance in the DEM group, indicating that both coherence and engagement of DMNs (microstate F) are altered. Strikingly, both spatial coherence and occurrence of microstate C distinguish NC, MCI, and DEM, with the lowest adjusted $p$-values given by statistical significance test in Table~\ref{tab:posthoc_results}. This confirms the SHAP ranking that placed \texttt{C\_mean\_corr} and \texttt{C\_occurrences} as the topmost impactful features, implicating progressive salience-network breakdown.

Taken together, these findings indicate that the classifier mainly detects a loss of salience/attention-network integrity (microstate C) while also relying on complementary disruptions of DMN activity (microstate F). It is consistent with these observations that neurodegenerative progression is accompanied by network hyperactivity (B/F) and functional breakdown (C), each with distinct temporal and structural signatures. Note that microstate F itself is a relatively new microstate linked to personally salient cognition, mental simulation, and theory-of-mind processes~\cite{brechet2019capturing}. Its position at the top of the SHAP ranking at the MCI stage (Figure \ref{fig:shap}(b)) suggests that anterior DMN is already compromised early in the disease course.

To determine whether microstate metrics differ across cognitive stages, we first assessed normality with the Shapiro–Wilk test; no feature met the assumption ($p>0.05$), so we applied the non-parametric Kruskal–Wallis H test~\cite{macfarland2016kruskal}.  As summarised in Table \ref{tab:stat_tests}, all features except \texttt{F\_mean\_corr} showed significant group effects ($p<0.05$). Dunn–Bonferroni post-hoc analysis (Table \ref{tab:posthoc_results}) confirmed that most features differed across every pair of groups, with the exceptions of \texttt{B\_mean\_corr}, \texttt{A\_mean\_corr}, and \texttt{B\_occurrences} for the NC–MCI comparison, and \texttt{B\_occurrences} and \texttt{A\_occurrences} in one dementia pairing—reflect the lower SHAP importance assigned to microstates A and B.

The significance test of the microstate features reiterate that temporal metrics (occurrences) are often more discriminative than spatial coherence (mean correlation), especially for microstate F. Crucially, only microstates C and F differentiate MCI from NC, identifying them as the earliest EEG markers, whereas alterations in microstates A and B emerge only at the DEM stage (See Figure~\ref{fig:microstate_features}).

These findings reinforce the hypothesis that both the temporal frequency and inter-state coherence of EEG microstates capture underlying neurophysiological differences between NC and DEM, which can be further studied in longitudinal experiments to identify neurological changes and disease progression more accurately. This suggests that microstate-based features can serve as clinically meaningful indicators of cognitive decline and may provide utility in early-stage DEM screening or disease progression monitoring.

\subsection{Model-Informed Feature Relevance}

SHAP-based global explanations from the best-performing EEG-MSAF-SVM classifier further validate the relevance of these microstate features. In the NC group, the model predominantly relied on correlation-based features (e.g., \texttt{C\_mean\_corr}, \texttt{F\_mean\_corr}). This indicates that, in healthy brains, both salience/attention (microstate C) and DMN (microstate F) networks exhibit high spatial consistency and sustained engagement.
For MCI, a shift was observed toward duration and mixed-metric features (e.g., \texttt{A\_occurences}, \texttt{B\_mean\_durs}), indicating early-stage compensatory dynamics, while microstate F coherence (\texttt{F\_mean\_corr}) enters the upper position of the SHAP ranking. This shift signals the first detectable disruption of anterior DMN integrity.

In the DEM group, we observed that \texttt{C\_mean\_corr} and \texttt{F\_mean\_corr} features are highly attributed. Then the SHAP scores are evenly distributed, placing higher importance on occurrence features (e.g., \texttt{F\_occurrences}, \texttt{C\_occurrences}), highlighting disrupted network regulation and diminished temporal stability. Notably, microstate C features—especially \texttt{C\_mean\_corr} and \texttt{C\_occurrences}—emerge as sensitive markers of decline, consistent with their salience-related functional roles.

\subsection{Microstate F as an Emerging Early Marker}

Our EEG-MSAF, interpretable-ML framework consistently ranks theta-band microstate F metrics (\texttt{F\_mean\_corr}, \texttt{F\_occurrences}, \texttt{F\_meandur})  among the most influential features for classifying both MCI and DEM. These results suggest that disruptions in anterior DMN coherence appear early in the disease course, echoing the reports of disengagement of the default mode network~\cite{celone2006alterations}. While replication in larger longitudinal samples is required, our study positions microstate F as a promising candidate biomarker and illustrates how explainable AI can uncover subtle neurophysiological signatures.

\subsection{Clinical and Methodological Implications}

These findings underscore the importance of interpretable microstate-derived features for capturing neurodegenerative dynamics. The combination of correlation and occurrence metrics characterizes the functional degradation versus compensatory reorganization. Furthermore, integrating SHAP-based explainability provides model transparency, ensuring that predictions are grounded in neurophysiologically meaningful signals. This approach bridges the gap between black-box ML and clinically actionable biomarkers, facilitating trust and adoption in real-world diagnostic settings.

\subsection{Limitations and Future Work}

While this study provides robust evidence for the utility of microstate features, it is limited to global summary statistics sample size, static modelling, and single-modality focus. Future work will explore subject-level SHAP distributions, temporal transitions between microstates, and multi-modal integration (e.g., with connectivity, bio-clinical data) to further enhance sensitivity and specificity. Additionally, expanding the dataset to include longitudinal MCI-to-AD converters would enable predictive modeling of disease progression.

\section{Conclusion}\label{conclusion}

In this study, we proposed an interpretable end-to-end framework for DEM classification using EEG microstate features. By leveraging microstate-derived metrics and classical ML models — specifically SVM, Random Forest, and XGB — we achieved state-of-the-art performance on two clinical EEG datasets. Notably, EEG-MSAF-SVM outperformed DL baselines (CEEDNET on CAUEEG and EEGConvNeXt on the Thessaloniki dataset), achieving 89\% and 95\%  classification accuracy, respectively, under 5-fold cross-validation.

Beyond predictive performance, our framework enables transparent model interpretability through SHAP-based feature importance, revealing that microstate correlation and occurrence metrics are key discriminators across disease stages. Our statistical and visual analyses further highlight systematic alterations in spatiotemporal microstate dynamics, particularly the reduced engagement of microstate C and increased compensatory shifts in microstate A/B/F, as markers of cognitive decline.

This work not only confirms the neurophysiological relevance of microstates in DEM but also emphasizes the feasibility of combining explainable AI with lightweight ML models for real-world clinical deployment. Future work may extend this framework to multi-modal integration and adaptive monitoring in longitudinal cohorts.

\bibliographystyle{unsrt}
\bibliography{ref}

\end{document}